\documentclass[12pt,epsf,amssymb,qsymbols]{article}
\usepackage{tabularx}
\usepackage{array}
\usepackage{graphics}
\usepackage{graphicx}
\usepackage{epsfig}
\usepackage{amsmath}
\usepackage{amssymb}
\makeatletter

\usepackage{verbatim}

\newcommand{\bea}{\begin{eqnarray}}
\newcommand{\eea}{\end{eqnarray}}
\newcommand{\beq}{\begin{equation}}
\newcommand{\eeq}{\end{equation}}
\newcommand{\gev}{{\rm GeV}}

\newcommand{\pdir}{p\kern -5.2pt\raise 0.2ex\hbox {/}}
\newcommand{\vdir}{v\kern -5.75pt\raise 0.15ex\hbox {/}}
\newcommand{\kdir}{k\kern -5.75pt\raise 0.15ex\hbox {/}}
\newcommand{\epsdir}{\epsilon\kern -5.0pt\raise 0.15ex\hbox {/}}
\newcommand{\bvdir}{\bar{v}\kern -5.75pt\raise 0.15ex\hbox {/}}
\newcommand{\Ddir}{D\kern -7.75pt\raise 0.20ex\hbox {/}}
\newcommand{\ldir}{l\kern -5.0pt\raise 0.2ex\hbox{/}}
\newcommand{\varepsdir}{\varepsilon\kern -5.5pt\raise 0.15ex\hbox{/}}

\makeatother

\renewcommand{\thefootnote}{\fnsymbol{footnote}}
\begin{document}
\thispagestyle{empty} 
\begin{flushright}
\begin{tabular}{l}
{\tt CPT-2002/P.4459}\\
{\tt LPT Orsay 02-112}\\
{\tt Roma-1348/02}
\end{tabular}
\end{flushright}
\begin{center}
\vskip 1.2cm\par
{\par\centering \LARGE \bf Possible explanation of the discrepancy of the light-cone QCD sum rule 
calculation of $g_{D^\ast D\pi}$ coupling with experiment}\\
\vskip 0.75cm\par
{\par\centering \large  
\sc D.~Becirevic~$^a$, J.~Charles~$^b$, 

A.~Le~Yaouanc~$^c$, L.~Oliver~$^c$, O.~P\`ene~$^c$
and J.C.~Raynal~$^c$}
{\par\centering \vskip 0.5 cm\par}
{\sl 
$^a$ Dip. di Fisica, Universit\`a di Roma ``La Sapienza",\\
Piazzale Aldo Moro 2, I-00185 Rome, Italy. \\                                   
\vspace{.25cm}
$^b$ Centre de Physique Th\'eorique, CNRS Luminy, Case 907~\footnote{Unite propre de 
Recherche du CNRS - UPR 7061}, \\
F-13288 Marseille Cedex 9, France.
\\                                   
\vspace{.25cm}
$^c$ Universit\'e de Paris Sud, Centre d'Orsay, LPT (B\^at.210)~\footnote{Unit\'e mixte de
Recherche du CNRS - UMR 8627.}, \\
F-91405 Orsay Cedex, France.}\\
 
{\vskip 0.25cm\par}
\end{center}

\begin{abstract}
The introduction of an explicit {\it negative} radial excitation contribution
in the hadronic side of the light cone QCD sum rule (LCSR) of Belyaev, Braun, Khodjamirian 
and R\"uckl, can explain the large experimental value of $g_{D^\ast D\pi}$, recently measured 
by CLEO. At the same time, it considerably improves the stability of the sum rule when varying the Borel 
parameter $M^2$.
\end{abstract}
\vskip 0.2cm
{\footnotesize {\bf PACS:} \sf 14.40.Lb (Charmed mesons), \ 13.75.Lb (Meson-meson interactions),
\ 11.55.Hx (Sum rules)}                 
\vskip 2.2 cm 
\setcounter{page}{1}
\setcounter{footnote}{0}
\setcounter{equation}{0}
\noindent

\renewcommand{\thefootnote}{\arabic{footnote}}
\vspace*{-1.5cm}

\setcounter{footnote}{0}

\setcounter{equation}{0}
\section{Recalling the difficulty}
The discrepancy of the LCSR prediction for 
$g_{D^\ast D \pi}$ with the recent experimental result by CLEO is 
a striking and puzzling one. We recall the reader that the LCSR prediction 
made in ref.~\cite{belyaev} reads $g_{D^\ast D \pi}=12.5$, which got even 
smaller after the radiative corrections have been included~\cite{khodjamirian}, namely 
$g_{D^\ast D \pi}=10.5 \pm 3.0$. On the other hand the experimental value is 
$g_{D^\ast D \pi}=17.9 \pm 0.3 \pm 1.9$~\cite{CLEO}. 
Thus the discrepancy is large~\footnote{Notice that we speak here of the 
discrepancy of central values. Of course, there are errors in the sum rule estimate, 
but it is difficult to draw strict conclusions from them, since their estimate is itself, 
of course, uncertain.}. The above LCSR result has been very 
carefully discussed, the calculation has been 
improved several times, and it has also been verified by using other sum rule 
techniques~\cite{colangelo}. Although the QCD sum rule approach certainly suffers 
from large uncertainties, in most cases a good agreement with experiment has been 
obtained. Therefore, we cannot be satisfied with concluding by a comfortable attitude 
of skepticism against the whole sum rule approach. Still more puzzling is the fact that 
this quantity does not seem a priori to have anything particularly exotic. 
In addition, the other theoretical approaches do not show such a discrepancy for this 
quantity. A careful discussion meant to reduce the uncertainties presented by quark models,
and performed in the framework of Dirac equation~\cite{dirac}, prior to the experimental 
measurement, has led to a result $g_{D^\ast D \pi} \simeq 18$. It should be stressed that this 
result has been obtained in the heavy quark limit (for the latest quark model discussion of 
$g_{D^\ast D \pi}$ see ref.~\cite{jaus}).  The recent (quenched) lattice QCD calculation has led to 
$g_{D^\ast D \pi} = 18.8 \pm 2.3^{+1.1}_{-2.0}$~\cite{herdoiza}. It is therefore important to 
understand the specific difficulty which the standard sum rule approach seems 
to encounter in this case~\cite{kim}. In ref.~\cite{martina-franca} it has been 
noted that the simple quark-hadron duality ansatz which works in the one-variable
dispersion relations  might be too crude for the double dispersion relation.

\section{An appealing solution: a large negative radial excitation contribution 
to the LCSR} 

In sum rule calculations it is usual to assume that the higher state contributions 
can be included in the perturbative estimate of a continuum~\footnote{Except for 
the critical case of quark masses in the pseudoscalar 
correlator method, or in general for some refined calculations like the ones discussed 
in ref.~\cite{jamin}.}. 
In other words, one does not include an {\it explicit} contribution of an isolated 
excited state. The first reason is probably pragmatical: it seems unnecessary to
recourse to large excitation contributions if the stability criterion can be
well satisfied without them. The other, more theoretical, reason is that the excitation 
contributions are exponentially suppressed with respect to the lowest state by 
the Borel procedure, so that one can sufficiently account for them by the rough 
procedure of the perturbative treatment of the continuum.
Indeed, in the LCSR approach of ref.~\cite{belyaev}, this point of view has been adopted. 
We suggest, instead, that this neglect of an explicit radial excitation contribution 
may be the origin of the above discrepancy between the LCSR prediction and the experimental 
value for $g_{D^\ast D \pi}$. Of course, we are aware that, as explained in Belyaev et al.~\cite{belyaev}, 
in their method such contributions are suppressed by the Borel exponential. Our claim is
that, in this particular case, the Borel suppression is not sufficient to 
allow to neglect them. The inclusion of the explicit radial excitation contribution to 
the hadronic side of the LCSR (often referred to as the left hand side --l.h.s.-- of the sum rule) 
also offers an appealing explanation for the failure of the LCSR prediction.

A first indication in favor of this new proposal is that, after including a radial 
excitation, the stability of the sum rule, under the variation of the Borel parameter $M^2$, 
is improved.  This criterion is however subject to an uncertainty since we do not really
know what is the accuracy of the calculation of the theoretical r.h.s.
It would be therefore good to present cross-checks of our assumption, which is what we 
will do in the next sections.

Before embarking on those issues, let us first explain why the assumption of a large 
radial excitation coupling to the sum rule, {\it with a negative sign} relative to the ground 
state contribution, seems so interesting.  Since the purpose of this letter is to communicate our 
proposal, and not to make the best numerics, we leave apart for the moment the radiative 
corrections and write the sum rule of Belyaev et al. as:
\bea
g_{D^\ast D \pi}= {f(M^2) \over f_D f_{D^\ast} } \;,
\eea
where the function $f(M^2)$ is the r.h.s. of eq.~(44) of ref.~\cite{belyaev}, namely
\bea
f(M^2)={ m_c^2 \over m_D^2  m_{D^\ast}} \ f_{\pi} \phi_{\pi}(1/2) \ M^2
\exp\left( {m_D^2+m_{D^\ast}^2 \over 2 M^2 }\right) \left[ 
e^{- m_c^2/M^2} - e^{- s_0/M^2}\right]+ \dots
\eea        
We do not write the higher twist terms since it would make the expression lengthy
and would not help understanding our proposal. Those terms are numerically very important and 
are indeed included in our calculation. It is then found that, within the Borel window  
$2~\gev^2<M^2<4~\gev^2$ determined by the standard criteria, the function $f(M^2)$ 
is monotonously decreasing, and the variation is as large as $20 \%
$. Therefore, there is no truly good plateau. The authors quote an average 
\bea
\left.\overline f(M^2)\right|_{2~\gev^2<M^2<4~\gev^2}=\ 0.51\pm 0.05~\gev^2\;,
\eea
which then yields the central value :
\bea
g_{D^\ast D \pi}\ =\ 12.5 \;,	
\eea
indeed much too low. Let us now introduce a radial excitation contribution 
to the hadronic l.h.s. of eq.~(44) of ref.~\cite{belyaev}, or equivalently write:
\bea \label{include1}
g_{D^\ast D \pi}={1\over  f_D f_{D^\ast} }  \left[ f(M^2) -  
R_{D^\prime} \exp\left( -{m_{D'}^2-m_{D}^2 \over 2 M^2}\right) - 
R_{D^{\ast \prime}} \exp\left(-{m_{D^{\ast \prime}}^2 - m_{D^\ast}^2 \over 2 M^2 }\right) \right]\;.
\eea
Note that we have two extra contributions: either the $D$ or the  $D^\ast$ is excited; they 
are denoted as $(D^\prime, D^\ast )$ and $(D,D^{\ast \prime})$, leading respectively to :
\bea \label{R'}
R_{D^\prime}=\left( {m_{D^\prime }\over m_D} \right)^2 f_{D^\prime}f_{D^\ast}g_{ D^\ast D^\prime \pi}\,,\quad
R_{D^{\ast \prime}}={m_{D^{\ast \prime}}\over m_{D^\ast}} f_{D}f_{D^{\ast \prime}} g_{D^{\ast \prime }D\pi}\,.
\eea
We assume that the higher $(D^\prime,D^{\ast \prime})$, $(D^{\prime \prime},D^\ast)$, \dots 
contributions are still included in the continuum part of the model. The completion of the procedure  
requires also a new value of $s_0$, since the lowest radial contributions are no more comprised 
in the continuum part, but are handled separately.
 
\begin{figure}
\vspace*{-0.8cm}
\begin{center}
\begin{tabular}{@{\hspace{-0.25cm}}c}
\epsfxsize12.2cm\epsffile{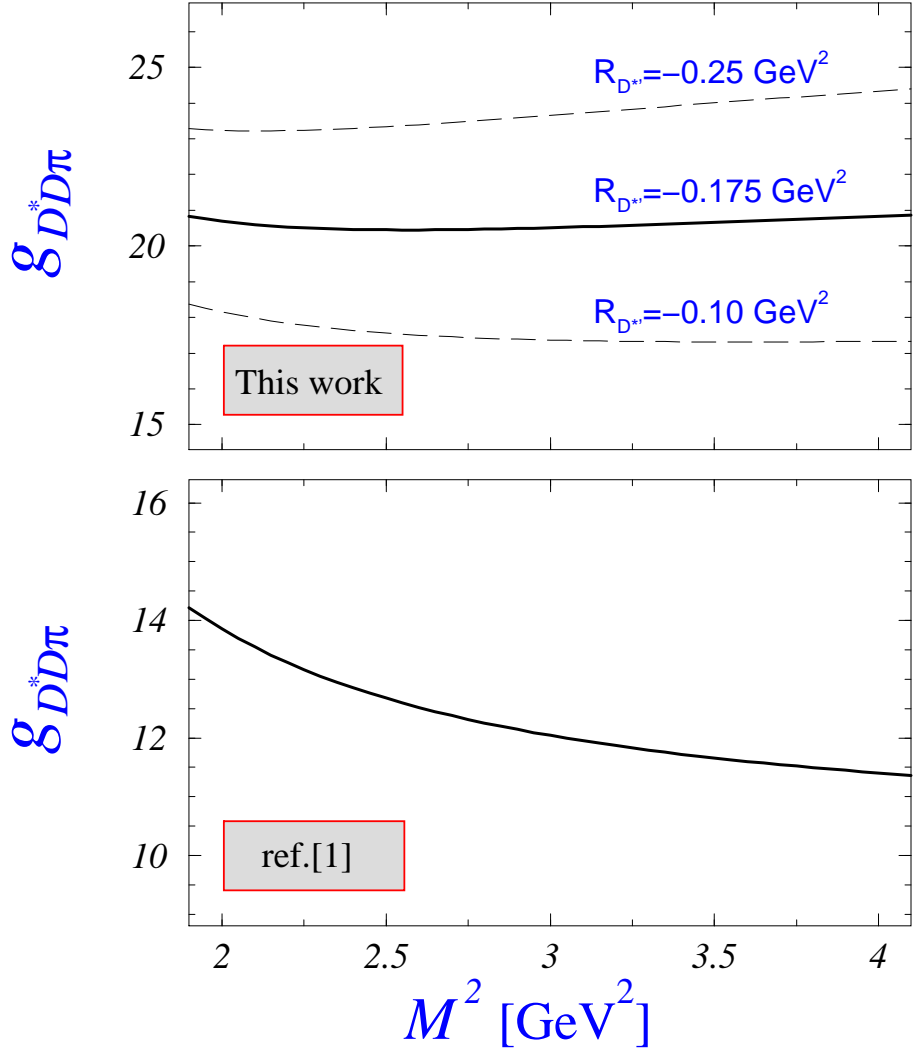}    \\
\end{tabular}
\caption{\label{fig:1}{\footnotesize The effect of inclusion of the radially excited $D^{(\ast) \prime}$-meson 
contributions to the hadronic part of the LCSR: In the lower figure the 
result of ref.~\cite{belyaev} is reproduced (no explicit radial excitations considered); In the 
upper curve we include the radial excitations as indicated in eq.~\eqref{include1} with the value of 
$R_{D^{\ast \prime}}$ varied in the range specified in~\eqref{range}. The latter range arises 
from the requirement of the stability of the sum rule~\eqref{include1} to be within the
$5$\% level, when the Borel parameter is varied as $2~ \gev^2 \leq M^2  \leq 4~ \gev^2$.} }
\end{center}
\end{figure}

For simplicity, since there is no sense in requiring too much precision, we will assume:
\bea
m_{D^\prime}=m_{D^{\ast \prime}}\,,\qquad f_{D^\prime}=f_{D^{\ast \prime}}\,,
\eea
the spin-spin effect being expected to decrease for higher states. We also 
assume that $g_{D^{\ast}D^\prime \pi}=g_{D^{\ast \prime }D\pi}=g'$, which is more questionable, 
since the ground state is present, and $D$  differ appreciably from $D^{*}$. This, however, 
should not alter the qualitative conclusion that we obtain. Note also that the 
difference between the $0^{-}$ and $1^{-}$ states would disappear in the heavy quark limit. 
With these assumptions, we have
\bea \label{ratio}
{R_{D^\prime}\over R_{D^{\ast \prime}}}\ =\ {m_{D^{\ast \prime}} m_{D^\ast}\over m_D^2 } {f_{D^\ast} \over f_{D}} \;.
\eea
This equation (\ref{ratio}) relates the two radial contributions since $R_{D^\prime}/R_{D^{\ast \prime}}$ 
is directly calculable if we know the mass $m_{D^{\ast \prime}}$. The exponentials accompanying 
$R_{D^\prime}/R_{D^{\ast \prime}}$ are the trace of the Borel suppression of radial excitations; 
they are {\it increasing} functions of $M^2$. 
Therefore, if we choose a {\it negative} $R_{D^\prime,D^{\ast \prime}}$, we can compensate 
for the decrease of $f(M^2)$ and thus simultaneously: (i) {\it improve the stability of the sum rule} and, 
(ii) {\it increase the magnitude of $g_{D^\ast D \pi}$ } (see fig.~\ref{fig:1} for illustration). 
We have of course to take a rather 
large radial contribution in order to get a significant effect. To avoid adding more freedom, 
we fix the excitation spectrum by a model calculation which is completely independent. 
Using Dirac equation, with numbers taken from our previous treatment~\cite{dirac}, we find 
in the heavy quark limit that the excitation energy is $0.5~\gev$, and therefore the mass 
of the excitation is $2.5~\gev$. Let us recall that the large width~\cite{melikhov} expected for such 
a state makes it difficult to observe.

The remarkable fact is the following: taking a negative $R_{D^{\ast \prime}}$ ($R_{D^\prime}$), 
and imposing that the variation of $g_{D^\ast D \pi}$ is $5\%
$ or less in the allowed window for $M^2$, (instead of the previous $20\%
$, in the same Borel window), we find that $R_{D^{\ast \prime}}$ must be in the range
\bea \label{range}
-0.25~\gev^2 < R_{D^{\ast \prime}} < -0.1~\gev^2\,. 
\eea
Then, depending on the value 
of $R_{D^{\ast \prime}}$ in the above range~\eqref{range} the predicted 
value of $g_{D^\ast D \pi}$ varies as
\bea
17\ <\ g_{D^\ast D \pi}\ < \ 25\;.
\eea
It increases with $\vert R_{D^{\ast \prime}} \vert $. It is much larger than before. 
We obtain at the same time a {\it much better stability in $M^2$}, and a much larger coupling 
constant, {\it just in the desired range to agree with experiment}.

For $R_{D^{\ast \prime}}=-0.15~\gev^2$, 
we would obtain 
a perfect stability in $M^2$ (accuracy of the order of $1\%
$) and the value $g_{D^\ast D \pi} = 19.4$, that is the value found in lattice QCD. 
But at the present stage, there is no sense in trying to get such an accuracy in 
stability, given the very rough precision of the calculation.

For the present note, we have adopted the parameters of ref.~\cite{belyaev} for the allowed range of variation $M^2$ ; we choose also to keep the continuum threshold to be the same as $s_0$, the continuum threshold of the two-points annihilation sum rules proposed by these authors, although this may be questioned since we now take apart certain contributions of excited states; we do not indulge in any fine tuning. We have also not allowed the 
annihilation constants $f_D$, $f_{D^{\ast}}$ to vary with $M^2$, but instead, we have taken 
the final value given in the same reference~\cite{belyaev}: indeed, we should check whether  
$f_D$ ,$f_{D^{*}}$ as well as $g_{D^\ast D \pi}$ are independent of $M^2$. 
As we stated above, the inclusion of the radial excitation stabilizes the LCSR. This does not
make the same effect in the sum rules for $f_D$, $f_{D^{*}}$. As we shall see, the radial 
excitation contribution is neither stabilizing the sum rule (w.r.t. the variation of the Borel 
parameter $M^2$), nor it creates a difficulty: it simply gives a negligible contribution, 
and does not modify the previously estimated values for $f_D,f_{D^{*}}$.

We note that the minus sign for the radial excitation contribution is crucial,
and we would like therefore to have some argument for this from models. 
There is no very trustable model for radial excitation strong decay at the quantitative level.
Nevertheless, it is indicative for our purpose that the rather standard model 
for strong interaction couplings, the non relativistic quark pair creation (QPC) model, 
used for calculational simplicity with harmonic oscillator wave functions, 
give precisely a stable {\it negative sign} for the product 
$f_D f_{D^{\ast \prime}} g_{D^{\ast \prime }D\pi}$ relative to the similar 
product for the ground state $f_D f_{D^{*}} g_{D^\ast D \pi}$. 
Hence, since, in the sum rule, 
the latter product is positive, this means that 
$R_{D^{\ast \prime}}$, which is precisely $f_D f_{D^{\ast \prime}} g_{D^{\ast \prime }D\pi}$ 
up to positive mass ratios, should be negative, as we have found in the sum rule.

\section{Check of $D \to \pi \ell \nu$ semileptonic decay} 

We note that exactly the same quantity $R_{D^{\ast \prime}}$ will appear 
in a $t$-channel analysis of $D \to \pi \ell \nu$ decay. Indeed, using an 
unsubtracted dispersion relation, as required for 
the form factor~\cite{burdman}, the addition of the residue 
of the pole corresponding to the radial excitation results in :
\bea
F_+ (q^2)\ =\ {1 \over 2 m_{D^{\ast}}} { f_{D^\ast} g_{D^\ast D\pi} \over  1 - q^2/m_{D^\ast}^2 }\ +\ 
{1 \over 2 f_D m_{D^{\ast \prime}}^2} { m_{D^\ast} R_{D^{\ast \prime}} \over  1 - q^2/m_{D^{\ast \prime}}^2 } \  +\  \dots
\eea
So the question is whether our ``large" contribution of radial states in 
the LCSR leads to effects in the $D \to \pi \ell \nu$ which are compatible 
with the facts. The answer is first that its impact is rather small, and anyway 
rather in the right direction. Indeed, we find that for $R_{D^{\ast\prime}}=- 0.15~\gev^2$, 
allowing $g_{D^\ast D \pi} \simeq 19$, the residue of the radial excitation
is almost one order of magnitude smaller than the ground state one :
\bea  
F_+(q^2)\ = \ {1.15 \over 1 - 0.25 q^2 }\ -\ {0.14 \over  1 - 0.16 q^2 }\  +\   \dots
\eea
It should be noted that the ground state contribution alone would lead 
to a {\it too large rate} (the numerator should be about $0.8$ at most instead of $1.15$ 
to get the correct rate with only the first pole). Therefore, the negative radial 
contribution is not worrisome but, quite the contrary, it is not sufficiently large to
balance the excess of the lowest state contribution. Other radial excitations, as well 
as the continuum, must contribute, which is in agreement with the conclusions of 
ref.~\cite{burdman}. Analogously, a negative contribution appears in the Becirevic--Kaidalov 
model~\cite{kaidalov}, from remote singularities.

\section{Check of annihilation constants and Adler--Weisberger sum rule}

Another worry could be that this large contribution could contradict 
other sum rules concerning either the annihilation constants or the 
strong couplings. The annihilation constants $f_D$ and $f_{D^\ast}$ 
are determined from the QCD sum rules by using the  $P$-$P$ and $V$-$V$ 
correlation functions, respectively. At the same time, 
an Adler--Weisberger sum rule constrains the strong interaction coupling 
constants. Since the latter is stronger, as we shall see, we begin with it and write it, in the case of $\pi +D \to \pi +D $ scattering, as :
\bea
\sum_{n=0}^{\infty} {f_{\pi}^2 \over 4 m_{D^{\ast (n)}}^2} \vert g_{D^{\ast(n)}D\pi} \vert^2 + 
\mbox{orbitally~excited~states}\ =\ 1\;,
\eea
where we have made explicit only the $D^{\ast}$ ground state contribution and the one of its {\it radial} 
excitations ; the allowed orbital excitations are the $0^{+}$ and $2^{+}$. In ref.~\cite{dirac}, we have estimated these contributions of the ground state 
and lowest orbitally excited states ($D^{\ast \ast}$) to be $70 \%
$, whence
\bea \label{gborne}
\vert g_{D^{\ast\prime}D\pi}\vert  {f_{\pi} \over 2 m_{D^{\ast\prime}}} < \sqrt{ 0.3 } = 0.55 \;.
\eea
Now, since we have  $\vert R_{D^{(\ast) \prime}} \vert  > 0.1~\gev^2 $, we can deduce a lower 
bound on $f_{D^{ \ast \prime}}=f_{D^\prime}$ from eq. (\ref{R'})
\bea
\vert f_{D'} \vert   \gtrsim 0.02\ \gev\;.
\eea 
This is one magnitude smaller than the ground state annihilation constant. 
Studying the QCD sum rules for $f_D$, $f_{D^\ast}$, we find that the resulting contribution
from the radially excited states is very small at the level of this lower bound.
For the pseudoscalar sum rule, one has to change the l.h.s. according to 
\bea
f_{D}^2 \to f_{D}^2 + \left({m_{D^\prime}\over m_D}\right)^4 f_{D^\prime}^2 
\exp\left(- {m_{D^\prime}^2-m_{D}^2 \over M^2}\right)\;.
\eea
Numerically, that change is completely negligible: it adds less than $0.0016~\gev^2$ 
(times the exponential factor, dependent on $M^2$, and which is smaller than $0.25$) 
to $f_D^2=0.17^2=0.029~\gev^2$.  Therefore, we have much room left even 
if the bound~(\ref{gborne}) were to be lowered. For the vector sum rule, 
the conclusion is quite similar. The substitution to be made is
\bea
f_{D^{\ast}}^2 \to f_{D^{\ast}}^2 + \left({m_{D^{\ast \prime }}\over m_D^{\ast}}\right)^2 
f_{D^{\ast \prime}}^2 \exp\left(- {m_{D^{\ast \prime}}^2-m_{D^{\ast}}^2 \over M^2} \right)\;.
\eea
The change is numerically even smaller in this case.

In contrast to the LCSR for $f_{D}f_{D^{\ast}} g_{D^\ast D \pi}$, 
the radial excitation term does not improve the stability of the sum 
rules for the annihilation constants: indeed, the r.h.s. is decreasing 
with $M^2$, and  the contribution from the radially excited state increases, 
since it is positive. But it does not deteriorate the stability, 
since it is very small.

\section{Conclusion}
The result of the proposed modification of the sum rule 
calculation is very encouraging. Introducing the radial excitation 
contribution significantly improves the value of $g_{D^\ast D \pi}$, and at the 
same time the stability of the sum rule with respect to the Borel parameter $M^2$. 
This means that the effect of such a radial state is not properly accounted by the 
standard perturbative continuum contribution. Unless one imposes an 
unreasonable degree of stability, one is not able to fix the magnitude 
of the radial contribution accurately. The latter may vary by more 
than a factor of two, which induces an important variation 
of $g_{D^\ast D \pi}$, around $30\%
$. This should not be viewed as a problem since our main goal is to 
prove that a large value for $g_{D^\ast D \pi}$, i.e. in the 
range allowed by experiment, is possible in the LCSR approach. At the 
very least, we can say that the old solution with the low value of 
$g_{D^\ast D \pi}$ and no explicit radial excitation, is not compelling, 
and alternative ones with large values of $g_{D^\ast D \pi}$ are favored. 
Of course, one should wonder whether the success survives when the 
calculation of the theoretical side is improved. In particular, does 
it survive the introduction of the radiative corrections, which are 
large? Our answer is positive, but we would like to present it within 
a more extensive discussion of the parameters involved in the sum rule 
calculation, especially because the threshold parameter $s_0$ should depend on 
the fact that we separate an explicit contribution of some excited states.

The main reason for having such a large effect of radial 
excitations without exceedingly large couplings is that the 
{\it Borel exponential suppression effect is very weak in the LCSR}: 
at most a factor $0.5$ (at the lower end of the allowed range) for the relative
suppression of the radial excitation with respect to the ground state. The Borel 
suppression is much less effective than in the standard sum rules, e.g. for  
correlators leading to $f_{D^{(\ast)}}$. This is due to the additional factor 
of ${1/2}$ in the exponent and to the fact that $M^2$ is twice larger 
on the mean in the allowed range ($2~\gev^2<M^2<4~\gev^2$, instead of 
$1~\gev^2<M^2<2~\gev^2$). At the lower end of the range, the suppression 
is around eight times less effective in the LCSR.

Of course, it would be good to examine whether one gets similar 
improvement for the other sum rule approaches to $g_{D^\ast D \pi}$. 
On the other hand, the situation encountered here underlines the 
usefulness of an alternative to the Borel transformation 
method, able to isolate with more efficiency various states. In some cases, 
the  FESR proposed for the case of the quark mass sum rules by Kambor
and Maltman~\cite{maltman} could be a good alternative. 

\section*{Acknowledgement} We thank Alexander Khodjamirian for discussions and 
precious comments on the manuscript. We acknowledge the partial support 
from the EC contract HPRN-CT-2002-00311 (EURIDICE).

\end{document}